\documentstyle[aps,preprint,version2]{revtex}   

\begin{document}
\draft

\begin{title}
Quantum fluctuations of $D_{5d}$ polarons on $C_{60}$
molecules
\end{title}
\author{Chui-Lin Wang}
\begin{instit}
China Center of Advanced Science and Technology
(World Laboratory), Beijing 100080, China
\end{instit}
\author{Wen-Zheng Wang, Yu-Liang Liu and Zhao-Bin Su}
\begin{instit}
Institute of Theoretical Physics,
Academia Sinica, Beijing 100080, China
\end{instit}
\author{Lu Yu}
\begin{instit}
Institute of Theoretical Physics,
Academia Sinica, Beijing 100080, China\\
and International Center for Theoretical Physics,
P. O. Box 586, Trieste 34100, Italy
\end{instit}

\begin{abstract}
The dynamic Jahn-Teller splitting of the six equivalent $D_{5d}$ polarons
due to quantum fluctuations is studied in the framework of
the Bogoliubov-de Gennes formalism.
The tunneling induced level splittings are determined
to be $^2 T_{1u} \bigoplus ^2 T_{2u}$ and $^1 A_g \bigoplus ^1 H_g$
for $C_{60}^{1-}$ and $C_{60}^{2-}$, respectively,
which should give rise to observable effects
in experiments.
\end{abstract}
\narrowtext

The synthesis of buckminsterfullerene \cite{1}
has materialized the truncated icosahedral carbon molecule,
possessing the highest point group symmetry, and exhibiting pronounced
Jahn-Teller (JT) effects \cite{2}
due to the electron-phonon coupling
inherent in the high symmetry.  These effects determine
many important
physical and chemical properties of this third carbon isomer.
 Optical observations of phonon progressions and vibronic
peak broadening in neutral $C_{60}$
in solution, matrix and solid films, as well as
in its anion and cation counterparts clearly show the effect of JT
distorted $C_{60}$ molecule \cite{5,6}.
On the other hand, the ESR lineshape also hints at
the dynamic nature of these JT distortions \cite{6,7}.

In order to study the JT
distorted structures of $C_{60}^{-}$ and $C_{60}^{2-}$,
various theoretical approaches  \cite{9,10} have been used
and quantum chemical CNDO/MINDO type calculations \cite{11}
have been performed.
Based
on group-theoretical and topological analysis,
Ceulemans $et~al.$ \cite {12} have proposed that
the  symmetry of the stationary ground state of $C_{60}^{-}$
and $C_{60}^{2-}$ is determined by
a weak 2nd-order coupling. Depending on the sign of the relevant
parameter, the symmetry of this state may be D$_{5d}$ or D$_{3d}$.
These two nearly degenerate configurations were also found in
the quantum chemical calculations \cite{11}.

Meanwhile, within
the Bogoliubov-de Gennes (BdeG) formalism, well-known in the literature of
conducting polymers \cite{13},
a simple SSH-like Hamiltonian is adopted to deal with
the multi-mode JT problem of the molecule and the $D_{5d}$ symmetry
structures of $C_{60}^{1-}$ and $C_{60}^{2-}$ have been obtained
 \cite{14,15}.
For this static JT problem, the dynamic matrix of the different symmetry
configurations has to be  positive-definite to guarantee the dynamical
 stability. It has been found that the D$_{3d}$ symmetry is
not dynamically stable  within the BdeG formalism \cite{15}.
The self-trapped excitons in neutral and charged $C_{60}$ molecules have
 also been
studied using this framework \cite{wzw}.
This self-consistent formalism is very efficient in describing
the static
JT distortions of this type of multi-electron,
multi-mode systems. However, such a
quasi-static approach is not sufficient. In fact, there are six
equivalent
$D_{5d}$ polarons obtained by breaking the $I_h$ symmetry. These six
polarons states are degenerate in energy and are connected by quantum
tunneling, which may give rise to splitting of the energy state and
restoring the full $I_h$ symmetry. In fact,
the binding energy of the $D_{5d}$ polaron is rather small
(of the order of $40$ meV as seen from Ref. \cite{15}
and the present calculation), which is comparable
with the
zero point energy for some of the JT active $h_g$ modes. Therefore, it is
important to consider the dynamic JT effects due to quantum fluctuations.
The genuine ground state should be an irreducible representation of the
full $I_h$ group and play a crucial role in related physical and chemical
processes.

In this paper,
we will first briefly review the earlier results of the
BdeG formalism, and describe the static solutions of the  JT problem
for $C_{60}^{1-}$ and $C_{60}^{2-}$ anions. Then the
first-order degenerate perturbation theory is adopted to determine the
effect of the quantum fluctuations on the static
configurations. The secular equations driving the tunneling splitting
are solved analytically. We find that the quantum tunneling indeed
 leads to
a further increase of the polaron binding  energy by $\sim 15$ meV
and the full $I_h$ symmetry is recovered.
Finally, the possible relations of our
results with experiments and some neglected effects are discussed.

In Ref. \cite{15}, we simulated dynamic evolutions for a SSH-like
system to find the stationary ground state  configurations
of $C_{60}$ and $C_{60}^{1-}$ molecules.
We found that the singlet stationary configuration of $C_{60}^{2-}$
also has $D_{5d}$
symmetry.
To study the effect of quantum fluctuations on the static JT
distorted $D_{5d}$ polaronic configuration,
we write down the full Hamiltonian in the second quantized form as

\begin{eqnarray}
\FL
H & = & H_e + H_{e{\rm -ph}} + H_{{\rm ph}} ~~,\\
&H_e & = -\displaystyle \sum_{<\!i,j\!>\,s} t_0
(c_{i\,s}^\dagger c_{j\,s}+ c_{j\,s}^\dagger c_{i\,s} )~~, \nonumber \\
& H_{e{\rm -ph}} &= \displaystyle \sum_{\mu} \alpha~
\sqrt{\hbar\over 2M\omega_\mu} F^\mu (\hat{b}_\mu +
\hat{b}_\mu^\dagger)~~,\nonumber \\
& H_{{\rm ph}} &= \displaystyle\sum_\mu \hbar\omega_\mu
(\hat{b}_\mu^\dagger \hat{b}_\mu + {1 \over 2})~~, \nonumber
\end{eqnarray}
where
\begin{equation}
F^\mu = \sum_{<\!ij\!>,s} (\vec{\xi}_i^\mu - \vec{\xi}_j^\mu)\cdot
{\vec{l_1}(ij) \over l_1 (ij) }
(c_{i\,s}^\dagger c_{j\,s}+ c_{j\,s}^\dagger c_{i\,s} )~~.
\end{equation}
Here $\alpha$ is the linear mode-independent vibronic coupling
constant, $\vec{\xi}_i^\mu$ is the amplitude of the normal  mode $\mu$
in the homogeneous ground state as evaluated in Ref. \cite{15},
$\vec{l_1}(ij)/l_1(ij)$ is the unit vector along the unrenormalized bond
$<i,j>$.
$\hat{b}_\mu(\hat{b}_\mu^\dagger)$ is the annihilation(creation) operator
of the $\mu$th normal phonon mode, defined as $\hat{b}_\mu|0\!> =0$,
in which
$|0\!>$ is the unrenormalized phonon vacuum.
Each of the six $D_{5d}$ vacua breaks the $I_h$ symmetry and can be
defined
by an origin-shifting operator $\hat{U}^\beta$ as $|\tilde{o}^\beta\!> =
\hat{U}^\beta | o\!>$, acting on the undistorted vacuum.
This origin shifting unitary
operator is constructed to translate the unrenormalized phonon vacuum
of lattice vibration to one of the six phonon vacua given by
$D_{5d}$ polaronic configurations $\{Q_\mu^\beta\}$, $\beta=1,
\ldots,6$,

\begin{equation}
\hat{U}^\beta = \exp \{ -{1\over \hbar} \sum_\mu
\sqrt{M\hbar\omega_\mu\over 2}(\hat{b}_\mu - \hat{b}_\mu^\dagger)
Q_\mu^\beta \}~~.
\end{equation}
Here $Q_\mu^\beta$'s are c-numbers determined self-consistently
through the gap equation
$Q_\mu^\beta = - {\alpha \over M \omega_\mu^2 }
<\! e_\beta | F^\mu | e_\beta\!>$,
where the multi-electron wavefunction $|e_\beta\!>$ is the solution of the
corresponding BdeG equation in the same $\{Q_\mu^\beta\}$ configuration.

The shifted phonon operator of
$\beta$th $D_{5d}$ polaron has the form of
${\hat{b}_\mu}^\beta = \hat{U}^\beta \hat{b}_\mu \hat{U}^{\beta\dagger}$,
and the corresponding phonon vacuum is defined as
${\hat{b}_\mu}^\beta | \tilde{o}_\beta \!>  = 0$,
where
\begin{equation}
| \tilde{o}_\beta\!>
= \exp ( \sum_\mu \sqrt{M\omega_\mu\over 2\hbar} \hat{b}_\mu^\dagger
Q_\mu^{\beta} ) |o\!>
\exp( -\sum_\mu {M\omega_\mu\over 4\hbar}
(Q_\mu^{\beta})^2 )
\end{equation}
which satisfies the  normalization condition,
$ <\! \tilde{o}_\beta | \tilde{o}_\beta \!> = 1 $.
$\hat{b}_\mu^{\beta\dagger}$ describes the phonon excitation on the
shifted
vacuum.
In the above description, we have already made an approximation,
assuming that the $D_{5d}$ polaron and the undimerized state have the same
phonon spectrum, which will not significantly affect our result.

The quantum tunneling between six minima is a tunneling process of the
self-consistent lattice plus $\pi$-electron states.
Taking the states at the six equivalent $D_{5d}$ polaronic minima of the
adiabatic potential as the zero-order degenerate states,
the state vector of the
${\beta}$th polaron can be decomposed into two parts
$|\Psi_\beta \! > = |e_\beta \!> \bigotimes |\tilde{o}_\beta \!>,~
\beta= 1,2,\dots,6$.
The tunneling state should be a combination of the six non-orthogonal
$D_{5d}$ polaron
multi-electron states with different phonon vacua
$|\Phi_i\!> = \sum_{\beta=1}^6  C_{\beta i}|\Psi_\beta\!>$,
and can be solved by the degenerate perturbation theory.
The secular equation is given by:
\begin{equation}
 \left\| <\!\Psi_{\beta'}| H |\Psi_\beta\!> -
E<\!\Psi_{\beta'}|\Psi_\beta\!> \right\| = 0~~,
\label{sec}
\end{equation}
where
\begin{equation}
<\!\Psi_{\beta'}|H|\Psi_\beta\!> =
<\!\Psi_{\beta'}|H_e|\Psi_\beta\!> +
<\!\Psi_{\beta'}|H_{e{\rm -ph}}|\Psi_\beta\!> +
<\!\Psi_{\beta'}|H_{{\rm ph}}|\Psi_\beta\!> ~~,\nonumber
\end{equation}
\begin{eqnarray}
<\!\Psi_{\beta'}|H_e|\Psi_\beta\!>
& =& -\displaystyle t_0 \sum_{<\!i,j\!>\,s}
<\! e_{\beta'}|
(c_{i\,s}^\dagger c_{j\,s}+ c_{j\,s}^\dagger c_{i\,s} ) |e_\beta\!>
<\!\tilde{o}_{\beta'}|\tilde{o}_\beta\!> ~~, \nonumber\\
<\!\Psi_{\beta'}|H_{e{\rm -ph}}|\Psi_\beta\!>
&=& \displaystyle \alpha \sum_{<\!i,j\!>\,s} ~
<\! e_{\beta'}| F^\mu |e_\beta\!>
{1\over 2}(Q_{\mu}^{{\beta'}} + Q_{\mu}^\beta )
<\!\tilde{o}_{\beta'}|\tilde{o}_\beta\!> ~~, \nonumber\\
<\!\Psi_{\beta'}|H_{{\rm ph}}|\Psi_\beta\!>
&=& \displaystyle <\! e_{\beta'}|e_\beta\!>
 \sum_{\mu} ( {M \omega_\mu \over 2} Q_{\mu}^{{\beta'}} Q_{\mu}^\beta
+ {\hbar \omega_\mu \over 2} )
<\!\tilde{o}_{\beta'}|\tilde{o}_\beta\!>~~. \nonumber
\end{eqnarray}
Here
\begin{equation}
<\!\tilde{o}_{\beta'}|\tilde{o}_\beta\!>
= \exp \{ - \sum_\mu {M\omega_\mu \over 4 \hbar} ( Q_\mu^{{\beta'}}
- Q_\mu^\beta)^2 \}~~,
\end{equation}
and the matrix elements of
$<\! e_{\beta'}|
(c_{i\,s}^\dagger c_{j\,s}+ c_{j\,s}^\dagger c_{i\,s} ) |e_\beta\!>$
and $<\! e_{\beta'}| e_\beta\!>$
are Slater determinants which can be calculated in accordance with the
formulas listed in the Appendix of Ref. \cite{clw}.
The overlap integral has the form of
$<\!\Psi_{\beta'} |\Psi_\beta\!> = <\!e_{\beta'}|e_\beta\!>
<\!\tilde{o}_{\beta'}|\tilde{o}_\beta\!>$.
It is worth pointing out that $<\!\Psi_{\beta'} |H|\Psi_\beta\!>=
<\!\Psi_\beta |H|\Psi_{\beta'}\!>$.

To explicitly obtain the matrix elements, we number the axis pointing
to the north pole as axis 1, while
the remaining five axes spanning a pentagon-solid angle
are numbered as 2 to 6, respectively. Due to symmetry consideration,
we only need to calculate three kinds of  matrix elements.
When ${\beta'}=\beta$, the diagonal matrix elements
$<\!\Psi_{\beta'} |H|\Psi_{\beta'}\!>$
$(\beta = 1,2,\cdots,6)$ are all equal, being the potential energy
of $D_{5d}$ polarons.  For the $C_{60}^{1-}$ problem,
the off-diagonal matrix elements only have
two different values depending on the angle between the two axes.
It is clear from the geometry that all the angles between two axes
have a value $\sim 63.4^\circ$ except  angles between axes 2 and 4,
2 and 5, 3 and 5, 3 and 6 and 4 and 6,
which have a value  $\sim 180^\circ-63.4^\circ$.
So it is only necessary to evaluate the elements
 $<\!\Psi_1|H|\Psi_2\!>$ and
$<\!\Psi_2|H|\Psi_4\!>$. Moreover, since the HOMO state
in $D_{5d}$ polaron
has an ungerade parity which is singly occupied,
the $D_{5d}$ polaron
has an ungerade parity as well.
 Since the angles
of these two pairs of axes are supplementary to each other,
$<\!\Psi_2|H|\Psi_4\!>= -<\!\Psi_1|H|\Psi_2\!>$.
The above analysis can
also be applied to evaluate $<\!\Psi_{\beta'}|\Psi_\beta\!>$. The
secular equation (\ref{sec}),  now becomes
\begin{equation}
\label{mat}
\left \| \begin{array}{rrrrrr}
         \sigma & \tau &
            \tau & \tau &
            \tau & \tau \\
         \tau & \sigma &
            \tau & -\tau &
            -\tau & \tau \\
         \tau &  \tau &
            \sigma & \tau &
            -\tau & -\tau \\
         \tau & -\tau &
            \tau & \sigma &
            \tau & -\tau \\
        \tau & -\tau &
            -\tau & \tau &
            \sigma  & \tau \\
        \tau & \tau &
            -\tau & -\tau &
          \tau &\sigma  \\
          \end{array} \right \| = 0
\end{equation}
where $\sigma = H_{11} - E$, $\tau = U - ES$,
and $H_{11}=<\!\Psi_1|H|\Psi_1\!>$, $U=<\!\Psi_1|H|\Psi_2\!>$
and\\
 $S=<\!\Psi_1|\Psi_2\!>$.
Eq. (\ref{mat}) can be analytically reduced to
\begin{equation}
\left ( ( H_{11} - E )^2 - 5(U-ES)^2 \right )^3 = 0
\end{equation}
It is easy to obtain two three-fold degenerate
eigenvalues,
\begin{equation}
E_{\mp} = { H_{11} \mp \sqrt{5} U \over 1 \mp \sqrt{5} S}.
\label{emp}
\end{equation}

The six degenerate $D_{5d}$ polaron states span a six-dimensional
 space
carrying a reducible representation of the $I_h$ group,
and the tunneling between
them will remove this degeneracy and reduce the six--dimensional space
to two three-fold irreducible representations, $T_{2u}
\bigoplus T_{1u}$,
where the higher $E_{-}$ is identified to be $T_{2u}$ levels
by calculating characters of the representation matrices,
while the lower $E_{+}$ corresponds to $T_{1u}$ levels.
Moreover,
the eigenvectors can also be determined analytically in terms of
 $C_{\beta i} $,
which are listed in Table 1,
where the six tunneling split states $|\Phi_i\!>$, $i=1,2,\ldots,6$, are
orthogonal to each other through the relations
$<\!\Phi_i|\Phi_j\!> = \sum_{{\beta'}\beta}C_{{\beta'} i}^* C_{\beta j}
<\!\Psi_{\beta'}|\Psi_\beta\!>$.
 From Eq. (\ref{emp}), we find the energy splitting,
$E_{-} - E_{+} = (S H_{11} - U) 2 \sqrt{5} / (1 - 5 S^2)$.
Analyzing this expression, we find that the factor $S H_{11} - U$
is actually
equal to $-<\!\Psi_2 | H_{int}| \Psi_1\!>$, where
$H_{int} = H -
H_0$ and $H_0$ is the zeroth-order Hamiltonian which has $|\Psi_1\!>$
or $|\Psi_2\!>$ as its lowest eigenstates.
$<\!\Psi_2 | H_{int}| \Psi_1\!>$ represents the hopping integral
between the two states, which should be proportional to the energy
 splitting.
Hence, we have
\begin{equation}
E_{-} - E_{+} = - 2 <\!\Psi_2| H_{int}| \Psi_1\!>
{\sqrt{5} \over 1 - 5 S^2}~~,
\end{equation}
where $- 2 <\!\Psi_2 | H_{int}| \Psi_1\!>$ is the energy
splitting for the two $D_{5d}$ polaron problem,
and $\sqrt{5} /(1 - 5 S^2)$ is an enhancement factor
due to the presence of 5 neighbours.
For the parameter set of $t_0 = 1.91\ eV, \alpha=5.0\ eV/\AA$, we got
$H_{11} = -206.66327\ eV$, $U= -63.9725\ eV$, $S=0.30949$,
$<\!\Psi_2 | H_{int}| \Psi_1\!> = -0.01166\ eV$,
 $E_{-} = -206.579\ eV$, $E_{+} = - 206.679\ eV$, and
an energy splitting $E_{-}-E_{+} = 0.100\ eV$.

We have also studied the $C_{60}^{2-}$ polaron problem.
The difference between a $C_{60}^{2-}$
$D_{5d}$ polaron state and a $C_{60}^{1-}$ $D_{5d}$ polaron state
is that the former has a gerade parity, where the singlet HOMO
electronic
state is doubly occupied,
whereas the latter has an ungerade parity. Therefore
$<\!\Psi_2|H|\Psi_4\!>= <\!\Psi_1|H|\Psi_2\!>$ for $C_{60}^{2-}$ polaron.
The six-dimensional reducible representation can be reduced to
$A_g \bigoplus H_g$ as a result of tunneling splitting.
The solutions of tunneling states and corresponding wavefunctions of
$C_{60}^{2-}$ are given in Table 2.
For the parameter set of $t_0 = 1.91\ eV, \alpha=5.0\ eV/\AA$, we got
$H_{11} = -206.26237\ eV$, $U= -9.0269\ eV$, and $S=0.043731$, which
 results
in $E_{-} = -206.255\ eV$, $E_{+} =  -206.290\ eV$,
and an energy splitting, $E_{-}-E_{+} = 0.035\  eV$.

According to the above results of the dynamic problem,
the ground vibronic states of $C_{60}^{1-}$ and
$C_{60}^{2-}$ anions are $^2T_{1u}$ and $^1A_{g}$ states, respectively,
which are the same type of
symmetries as the unrenormalized electronic states in the
corresponding highest
symmetric configurations of $C_{60}^{1-}$ and $C_{60}^{2-}$.
These results are consistent with the expected
restoration of the initial $I_h$ symmetry through the quantum tunneling.
In principle, the vibronic reduction
effect in these vibronic ground states is observable, especially in the
ESR measurements. The estimate of Kato $et~al.$\cite{7} shows
that the observed $0.003$ reduction of {\tt g} factor cannot be
explained without the inclusion of JT effect. However, the combined
effects of JT vibronic and spin-orbit couplings need further
theoretical and experimental investigations. Furthermore, the mixing of
the vibronic ground states $^2T_{1u}$ and $^1A_g$ with the low-lying
excited $^2T_{2u}$ and $^1H_{g}$ states in the magnetic
field should also be taken into account to determine the above
observables. On the other hand, since the JT active modes $h_g$ are not
dipole-active, the IR transition between the vibronic states  $^2 T_{1u}$
and $^2 T_{2u}$, $^1 A_g$ and $^1 H_g$ are parity forbidden. However,
the tunneling splitting may be observable in the electronic
resonant Raman
scattering measurements, if the degenerate low-lying excited $^2
T_{2u}$ and $^1 H_g$ states are taken as the intermediate states.

Within our considerations, the pathways on the adiabatic potential
surface from one $D_{5d}$ minimum to another have not been determined
explicitly due to  the complications resulting from the
multi-mode nature
of the problem.
Based  on the work in the
$T_{1u} \bigotimes (e_g + t_{2g})$ problem \cite{16}, Khlopin $et~al.$
\cite{9} have shown that for the reduced linear one-mode
$T_{1u}\bigotimes h_g$ problem, the potential surface possesses a $2D$
equipotential continuum of minima in the strong coupling limit, and
indicated that the dynamic properties cannot be solved by
neglecting the multimode nature via a simple scale
transformation  because of the dynamic interaction between the
scaled coordinates. On the other hand, the locations and heights of the
possible $D_{3d}$, $D_{2h}$ and $C_{2h}$ saddle points on the potential
surface, corresponding to the lower ranking epikernels of $I_h$, are
the key factors to set the possible pathways. However, unfortunately,
the knowledge of  the height of these saddle points is not
sufficient to fully determine the pathways, because
the quadratic vibronic coupling is absent in our present approach.

Another neglected factor in the above discussions is the effect of
electron-electron correlations, which could affect both static and
dynamic aspects of the problem through the competition with
 the vibronic coupling \cite{kiv},
especially, in the case of multiply-charged anions.
However, the evidence of the important role of vibronic coupling
in the the dynamic process of the optical excitation experiments
is clear enough to show that the lattice of $C_{60}$ is
dynamically distorted and the symmetry is well defined
(not as $D_{5d}$)\cite{5,6,7}.  The dynamical
JT effects of charged fullerenes
have recently been discussed within a very different
approach (Berry phase)\cite{tos}. The large
tunneling splitting obtained in this calculation
supports the assumption of spherical symmetry, adopted there.

In summary, the static linear $T_{1u}\bigotimes 8 h_g$ JT effect is
analyzed for the $C_{60}^{1-}$ and $C_{60}^{2-}$ anions based on the
numerical results within the BdeG formalism. The dynamic problem has been
solved analytically within the standard perturbation approach. The
quantum tunneling among the six stationary equivalent $D_{5d}$
polaronic minima results in the  $^2 T_{1u}$ and $^1 A_g$
vibronic ground states, accompanied by the low-lying $^2
T_{2u}$ and $^1 H_g$ vibronic states for $C_{60}^{1-}$ and $C_{60}^{2-}$
anions, respectively. Therefore, in experiments one should look
for signatures of these states rather than the $D_{5d}$ states
obtained by neglecting the dynamical JT effects.

\vspace{1cm}
\noindent{\bf ACKNOWLEDGEMENTS}

The authors would like to thank W. M. You and E. Tosatti
for helpful discussions. This work is partially supported by
the National Natural Science Foundation of China.

\newpage

\squeezetable
\begin{tabular}{ccc}
\multicolumn{3}{l}{Table 1, Energies and Wave Function Coefficients
$C_{\beta i}$ of Tunneling states in $C_{60}^{-}$}\\
\hline
 & \hbox to 3truecm{} & \\
Sym. & E &
\multicolumn{1}{c}{\parbox[b]{6cm}{Wave function Coefficients \\
$N (C_{1 i}, C_{2 i}, C_{3 i}, C_{4 i}, C_{5 i}, C_{6 i})$}} \\
\hline
\\
$T_{2u}$
&
${ H_{11} - \sqrt{5} U \over 1 - \sqrt{5} S}$
&
$
\left\{ \begin{array}{l}
  (10(1-\sqrt{5}S))^{-1/2}(\sqrt{5},~~-1,
         ~~-1,~~-1,~~-1,~~-1)\\
  (40(1-\sqrt{5}S))^{-1/2}(0,4,
         -1-\sqrt{5},-1+\sqrt{5},
         -1+\sqrt{5},-1-\sqrt{5})\\
 (4(5-\sqrt{5})(1-\sqrt{5}S))^{-1/2}(0,0,
         -1+\sqrt{5},-2,2,1-\sqrt{5})
        \end{array} \right. $
\\
\\
\hline
\\
$T_{1u}$
&
${ H_{11} + \sqrt{5} U \over 1 + \sqrt{5} S}$
&
$
\left\{ \begin{array}{l}
(10(1+\sqrt{5}S))^{-1/2}(\sqrt{5},~~~1,
         ~~~1,~~~1,~~~1,~~~1)\\
(40(1+\sqrt{5}S))^{-1/2}(0,4,
         -1+\sqrt{5},-1-\sqrt{5},
         -1-\sqrt{5},-1+\sqrt{5})\\
 (4(5+\sqrt{5})(1+\sqrt{5}S))^{-1/2}(0,0,
         1+\sqrt{5},2,-2,-1-\sqrt{5})
        \end{array} \right. $
\\
\\
\hline
\end{tabular}
\vspace{1cm}

\squeezetable
\begin{tabular}{ccc}
\multicolumn{3}{l}{Table 2, Energies and Wave Function Coefficients
$C_\beta$ of Tunneling states in $C_{60}^{2-}$}\\
\hline
 & \hbox to 1truecm{} & \\
Sym. & E &
\multicolumn{1}{c}{\parbox[b]{6cm}{Wave function Coefficients \\
$N (C_1, C_2, C_3, C_4, C_5, C_6)$}} \\
\hline
\\
$H_{g}$
&
$ { H_{11} -  U \over 1 -  S}$
&
$
\left\{ \begin{array}{l}
(30(1-S))^{-1/2}(5,~-1,
                     ~-1,~-1,~-1,~-1)\\
(20(1-S))^{-1/2}(0,~~4,
                     ~-1,~-1,~-1,~-1)\\
(12(1-S))^{-1/2}(0,~~0,~~3,
                     ~-1,~-1,~-1)\\
(6(1-S))^{-1/2}(0,~~0,~~0,
                     ~~2,~-1,~-1)\\
(2(1-S))^{-1/2}(0,~~0,~~0,
                     ~~0,~~1,~-1)\\
        \end{array} \right. $
\\
\\
\hline
\\
$A_{g}$
&
$ { H_{11} + 5 U \over 1 + 5 S}$
&
$
   (6(1+5S))^{-1/2}(1,~~1,~~1,~~1,~~1,~~1)  $
\\
\\
\hline
\end{tabular}



\begin{references}
\bibitem{1} H. W. Kroto $et~al.$, Nature (London), {\bf 318}, 162(1985);
W. Kr\"{a}tschmer $et~al.$, Nature (London), {\bf 347}, 354(1990).
\bibitem{2} H. A. Jahn and E. Teller, Proc. R. Soc. (London), Sec A
{\bf 161}, 220(1937).
\bibitem{5} J. W. Arbogast $et~al.$, J. Phys. Chem., {\bf 95}, 11(1991);
C. Reber $et~al.$, {\it ibid}, {\bf 95}, 2127(1991);
Y. Wang, {\it ibid}, {\bf 96} 764(1992);
A. Skumanich, Chem. Phys. Lett., {\bf 182}, 486(1991);
Z. Gasyna $et~al.$, {\it ibid}, {\bf 183}, 283(1991);
J. P. Lane $et~al.$, Phys. Rev. Lett., {\bf 68}, 887(1992);
M. Matus $et~al.$, Phys. Rev. Lett. {\bf 68}, 2822(1992).
\bibitem{6} D. Dubois and K. M. Kadish, J. Am. Chem. Soc., {\bf 113},
4346(1991);
M. A. Greaney $et~al.$, J. Phys. Chem., {\bf 95}, 7142(1991);
Z. Gasyna $et~al.$, {\it ibid}, {\bf 96}, 1525(1992);
T. Kato $et~al.$, Chem. Phys. Lett., {\bf 180}, 446, {\bf 186}, 35(1991).
\bibitem{7} T. Kato $et~al.$, Chem. Phys. Lett. {\bf 205}, 405(1993).
\bibitem{9} V. P. Khlopin, V. Z. Polinger and I. B. Berzuker, Theor.
Chim. Acta (Beil.) {\bf 48}, 87(1978).
\bibitem{10} M. Lannoo $et~al.$, Phys. Rev. {\bf B44}, 12106(1991).
\bibitem{11} N. Koga and K. Morokuma, Chem. Phys. Lett. {\bf 196},
191(1992);
K.Tanaka $et~al$, $ibid$, {\bf 193}, 101 (1992).
\bibitem{12} A. Ceulemans, D. Beyens and L. G. Vanquickenborne,
J. Am. Chem. Soc., {\bf 106}, 5824(1984); A. Ceulemans, J. Chem. Phys.,
{\bf 84}, 6442(1986); {\it ibid}, {\bf 87}, 5374(1987); A. Ceulemans and
P. W. Fowler, Phys. Rev. {\bf A 39}, 481(1989); A. Ceulemans and
P. W. Fowler, J. Chem. Phys. {\bf 93}, 1221(1990);
A. Ceulemans  $et~al$, Structure and Bonding, {\bf 71}, 125 (1989)
\bibitem{13} L. Yu, "{\sl Solitons and Polarons in Conducting Polymers}",
(World Scientific, Singapore, 1988); "{\sl Conjugated Conducting
Polymers}", H. Kiess, ed., (Springer, Berlin, 1992).

\newpage

\bibitem{14} K. Harigaya, J. Phys. Soc. Japan, {\bf B60}, 400(1991);
Phys. Rev. {\bf B45}, 13676(1992); B. Friedman, Phys. Rev. {\bf B45},
1454(1992).
\bibitem{15} W. M. You, C. L. Wang, F. C. Zhang and Z. B. Su,
Phys. Rev. {\bf B47}, 4765 (1993);
W. M. You, C. L. Wang and Z. B. Su, preprint (1993).
\bibitem{wzw} W. Z. Wang, C. L. Wang, Z. B. Su and L. Yu,
submitted to PRL.
\bibitem{clw} C. L. Wang, W. Z. Wang, G. L. Gu, Z. B. Su and L. Yu,
Phys. Rev. {\bf B48}, 10788 (1993).
\bibitem{16} M. C. M. O'Brien, Phys. Rev. {\bf 187}, 407(1969).
\bibitem{kiv} S. Chakravarty, M. P. Gelfand and S. Kivelson,
Science {\bf 254}, 970 (1991);
B. Friedman and J. Kim, Phys. Rev.
{\bf B 46}, 8638(1992).
\bibitem{tos} A. Auerbach, N. Manini, and E. Tosatti; N. Manini, E.
Tosatti, and A. Auerbach, submitted to PR B.
\end{references}
\end{document}